\documentclass[submission,copyright,creativecommons]{eptcs}
\providecommand{\event}{Cassting 2016} 

\usepackage[english]{babel}
\usepackage[utf8]{inputenc}
\usepackage{amssymb}
\usepackage{amsthm}
\usepackage{amsmath}
\usepackage{tikz}

\usetikzlibrary{calc}
\usetikzlibrary{decorations.pathreplacing}

\newtheorem{theorem}{Theorem}
\newtheorem{lemma}{Lemma}

\newtheorem{proposition}{Proposition}

\newtheorem{remark}{Remark}

\newtheorem{definition}{Definition}

\newcommand{\nats}{\mathbb{N}}
\newcommand{\structnats}{\mathcal{N}}
\newcommand{\aut}[1]{\mathcal{{#1}}}

\title{Playing Games in the Baire Space}
\author{Benedikt Brütsch \quad\qquad Wolfgang Thomas
\institute{RWTH Aachen University}
\email{bruetsch@automata.rwth-aachen.de \quad\qquad thomas@automata.rwth-aachen.de\quad}
}
\def\titlerunning{Playing Games in the Baire Space}
\def\authorrunning{B. Brütsch \& W. Thomas}
\begin{document}
\maketitle

\begin{abstract}
We solve a generalized version of Church's Synthesis Problem where
a play is given by a sequence of natural numbers rather than a sequence of 
bits; so a play is an element of the Baire space rather than of the Cantor space. 
Two players Input and Output choose natural numbers in alternation to generate a play. 
We present a natural model of automata (``$\nats$-memory automata'')  equipped with 
the parity acceptance condition, and we introduce also the corresponding 
model of  ``$\nats$-memory transducers''. We show that solvability of games 
specified by $\nats$-memory automata (i.e., existence of a winning strategy for 
player Output) is decidable, and that in this case 
an $\nats$-memory transducer can be constructed that implements a winning strategy for 
player Output.
\end{abstract}


\section{Introduction}

The algorithmic theory of infinite games was started in 1957 when Church 
formulated his  ``synthesis problem''. This problem asked for presentation of a 
transformation $\alpha \mapsto \beta$ of $\omega$-sequences over a finite alphabet $\Sigma$, 
computable letter-to-letter and  satisfying a  
logical specification $R(\alpha, \beta)$. If we write $\alpha = \alpha(0) \alpha(1) \ldots$, 
$\beta = \beta(0) \beta(1) \ldots$ and set $\alpha \,\hat{\,}\, \beta =  \alpha(0) \beta(0) \alpha(1) \beta(1) \ldots $, 
the specification $R$ can be captured by the $\omega$-language 
$$L = \{\alpha \,\hat{\,}\, \beta \in \Sigma^\omega \mid R(\alpha, \beta)\}$$
Church's synthesis roblem asks: Given an
 $\omega$-language $L$ defined in a ``logistic system'', is there a letter-to-letter transformation 
$\alpha \mapsto \beta$ in the format of some kind of circuit such that $\alpha \,\hat{\,}\, \beta \in L$ for each $\alpha \in \Sigma^\omega$? 

In descriptive set theory a related question had been studied in game-theoretic 
terminology, regarding games between two players called here Input and Output, where $L$ serves as 
the winning condition for Output. A play  $\alpha(0) \beta(0) \alpha(1) \beta(1) \ldots$ 
is won by player Output if it belongs to $L$, and a transformation as mentioned above is then 
a winning strategy for Output. These ``Gale-Stewart games''  \cite{GaleStewart53} were studied 
in descriptive set theory focussing on the problem of determinacy  (whether 
one of the two players has a winning strategy). A major result in this theory says that if the 
set $L$ is Borel then the associated Gale-Stewart game, which we denote by $\Gamma(L)$,  is determined \cite{Kechris, Moschovakis}.
Church's Problem posed a sharpened question, namely, given a finite description of 
$L$, to determine who wins and to exhibit a concrete presentation of a winning strategy
for the winner. 

This problem was solved by B\"uchi and Landweber \cite{BL69} in the following strong sense:
If $L$ is a regular $\omega$-language (presented, e.g., by a deterministic Muller automaton), then 
the winner of the game $\Gamma(L)$ can be computed, and a winning strategy can be presented 
in the format of a finite-state machine (a Mealy automaton).  

This fundamental result has been extended in many ways, among them into the framework of 
infinite-state systems. For example, Walukiewicz  \cite{Walpushdown} showed the analogue of the 
B\"uchi-Landweber Theorem for pushdown systems.
It is remarkable, however, that a different kind of ``infinite extension'' of 
the B\"uchi-Landweber Theorem has not been addressed in the literature, namely the case where 
the input alphabet over which $\omega$-sequences are formed is infinite. 
Taking the typical case of a finite alphabet to be $2 = \{0,1\}$  
and the typical case of an infinite alphabet to be $\nats$ (we will focus solely on the alphabet $\nats$ 
in this paper), we 
are no more dealing with sequences from $2^\nats$ but from $\nats^\nats$. 

In set-theoretic topology (and in descriptive set theory) this is the step 
from the Cantor space  $2^\nats$ to the Baire space $\nats^\nats$. 
The topological classification theory of sets $L \subseteq \nats^\nats$ is 
developed in very close analogy to that of sets $L \subseteq 2^\nats$ (cf. \cite{Kechris, Moschovakis}), 
and determinacy of Borel games then holds for $\nats^\nats$ as it does for $2^\nats$. 
A small difference occurs in the representation of projective sets: In the 
Baire space, these can be described as projections of closed sets, whereas in 
the Cantor space one has to resort to projections of $G_\delta$-sets. 

In automata theory, however, the step to infinite alphabets is highly non-trival. It requires 
automata that work over infinite alphabets, in particular $\nats$.
Several proposals exist to introduce finite-state devices that can process 
finite or infinite words over an alphabet such as $\nats$. Let us recall some of them.

A straightforward approach is to code a number $m$ by a word over $\{0,1\}$, such as $0^m$ or $1^m$ or the 
binary expansion of $m$. Then we consider Banach-Mazur games in which 
a move by a player consists of a choice of a sequence of letters (from a finite 
alphabet) rather than single letters (for a recent reference on Banch-Mazur games see, e.g., \cite{GrLe12}). So a play 
				$m_0 m_1 m_2 \ldots$  may be coded by the bit sequence
				$0^{m_0+1} 1^{m_1+1} 0^{m_2+1} \ldots$ 
				in which a player contributes a word of $0^+$ or $1^+$. 
	A disadvantage of this approach is the fact that finite-state devices cannot check simple properties 
of ($\omega$-)words, for instance the equality of successive numbers of a play.
		
As models of automata working directly on infinite alphabets we mention the 
register automata of Kaminski and Francez \cite{KF}, the data automata 
of Bojanczyk et al. \cite{BDMS}, and the register automata over data words of \cite{DemriLazic} 
 that allow equality tests between 
letters (e.g., natural numbers) that occur in a word. Taking $\nats$ as the alphabet, 
a weakness of these models is their inability to check the order between successive 
letters or just the condition that a letter $m$ is followed by $m+1$ or by $m-1$. 

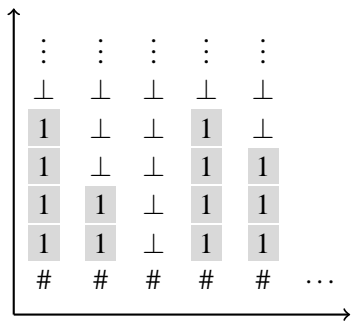
\begin{figure}[ht]\label{fig:grid}
	\centering
	
	\newcommand{\one}{\colorbox{black!15}{1}}
	\begin{tikzpicture}
		\node[inner sep=0.5em] (cols) {
			$\begin{matrix}
				\vdots & \vdots & \vdots & \vdots & \vdots & \\
				\bot   & \bot   & \bot   & \bot   & \bot   & \\
				\one   & \bot   & \bot   & \one   & \bot   & \\
				\one   & \bot   & \bot   & \one   & \one   & \\
				\one   & \one   & \bot   & \one   & \one   & \\
				\one   & \one   & \bot   & \one   & \one   & \\
				\#     & \#     & \#     & \#     & \#     & \cdots \\
			\end{matrix}$
		};
		
		\draw[thick] (cols.south west) edge[->] (cols.north west);
		\draw[thick] (cols.south west) edge[->] (cols.south east);
	\end{tikzpicture}
	
	\caption{Grid representation of the sequence of letters $4\,2\,0\,4\,3\,\ldots$}
\end{figure}

In the present paper, we work with automata which can check such relations of 
``incremental change'' between 
letters from $\nats$, called progressive grid walking automata (PGAs) and introduced recently 
in \cite{CST}. They cover all properties of the Banach-Mazur coding of sequences over $\nats$,
and they allow to check the relation between successive letters (as far as expressible in 
monadic second-order logic MSO over $(\nats, +1, 0)$). 
The idea is to code a letter by a column labelling of a labelled two-dimensional 
grid. A word $m_1 \ldots m_\ell$ is coded by a grid with $\ell$ column $\omega$-words, where the 
value $m_i$ is coded by the $\omega$-word $\# 1^{m_i} \bot^\omega$ as shown in Figure
\ref{fig:grid}. For $\omega$-words this grid is also right-infinite.

A progressive grid automaton is a three-way automaton that walks through such a grid 
from left to right, scanning a 
column in two-way mode and moving from a column to the next by a step to the right.
The latter feature allows to preserve a natural number value (the value $k$ if the step
to the right occurs at height $k$ of a column). This feature amounts to a memory for 
values in $\nats$. For example the value of an input $m_i$ may be handed from the $i$-th  
column to the next, by stepping out of the column at height $m_i$, in order to check, for example,  
that $m_{i+1} = m_i +1$. In the present paper we introduce a 
slightly stronger variant of PGAs, called \emph{$\nats$-memory automata}, and we also define a 
corresponding model of transducer. These automata use tokens of three kinds, ``memory token'', 
``memory update token'', and (for transducers) ``output token'' to indicate values of $\nats$. 
A column is scanned 
again in two-way mode, but starting from the bottom, using the memory token that is
located somewhere on the column (namely at the position where the memory update token was placed in the 
previous column), and in the current column the memory update token is then placed for handing 
a value $k$ of $\nats$ over to the next column (where the memory token will 
be on this position $k$). The transducer's output token is used to specify 
a value from $\nats$ as the result of a computation. 

Our main result will be an analogue of the B\"uchi-Landweber Theorem for $\nats$-memory automata:
\emph{Given a Baire space game $\Gamma(L)$ where $ L \subseteq \nats^\omega$ is defined 
by an $\nats$-memory 
automaton with parity acceptance condition one can decide who wins and construct 
an $\nats$-memory transducer that executes a winning strategy for the winner.}

The remainder of the paper is structured as follows: In the subsequent section we present 
some prerequisites on MSO-logic. In Section \ref{sec-automata} we introduce 
$\nats$-memory automata, first as acceptors of $\omega$-sequences over $\nats$, and then 
as transducers. In Section \ref{solving-games} we state and prove the main result. 
In the conclusion we address some 
perspectives and open problems.


\section{Prerequisites on MSO-Logic}\label{mso-prerequisites}


We assume that the reader is familiar with the basics on MSO-logic (as presented, e.g., in \cite{Tho97}). We recall known results 
to be used in later sections. 

It will be convenient to work with relational structures only. So we consider the structure 
${\cal N} = (\nats, {Succ})$ with the successor relation ${Succ}$ over $\nats$ rather than 
the structure $(\nats, +1, 0)$.

The MSO-theory of ${\cal N}$ is the set of all MSO-sentences that are true in ${\cal N}$. From 
\cite{Bu62} we know that this theory is decidable. 

We use two transfer results on preservation of the decidability of MSO-theories. The first 
refers to ``MSO-interpretations''. A structure ${\cal A} = (A, R^A)$, say with just one binary relation $R^A \subseteq A \times A$
is MSO-interpretable in a structure ${\cal B}$ (possibly with different signature)  if MSO-formulas $\varphi(x), \psi(x,y)$ exist that 
describe the structure ${\cal A}$ in ${\cal B}$, in the sense that the elements satisfying $\varphi(x)$ in ${\cal B}$
provide a copy of the domain $A$ in $B$, and the pairs $(a,b)$ satisfying $\psi(x,y)$ in ${\cal B}$ give 
a copy of $R^A$ over the copy of $A$. The following is well-known (see, e.g., \cite{EFT94}): 

\begin{proposition}
If the structure ${\cal A}$ is MSO-interpretable in the structure ${\cal B}$ and the MSO-theory of ${\cal B}$ is decidable, 
then so is the MSO-theory of ${\cal A}$.
\end{proposition}

The second model transformation is the step from a structure ${\cal A}$ to a product $[1, \ldots, k] \times {\cal A}$, 
the $k$-fold copy of $\mathcal{A}$. 
Let us consider just the case where ${\cal A} = {\cal N} = (\nats, {Succ})$. The domain of this product 
is the set $\{1, \ldots, k\} \times \nats$, and we have the following relations:
\begin{itemize}
	\item ${SUCC} = \left\{ \bigl( (r,n),(r,n+1) \bigr)  \mid  r \in [1, \ldots, k],\; n \in \nats \right\}$ 	
	\item $P_r = \left\{ (r,n)  \mid  n \in \nats \right\}$ (fixing membership in the $r$-th copy of ${\cal N}$)
	\item $\mathit{SameNumber} = \left\{ \bigl( (r,n),(s,n) \bigr)  \mid  r,s \in [1, \ldots, k],\; n \in \nats \right\}$
\end{itemize}
In the general case of a relational structure ${\cal A}$, the first item is applied to all relations that are present in ${\cal A}$.
It is easy to show (see, e.g., \cite{BCL07}) that the MSO-theory of $[1, \ldots, k] \times {\cal A}  $  is decidable
if the MSO-theory of ${\cal A}$ is decidable. We need here only the case ${\cal A} = {\cal N}$:     

\begin{proposition}
	For any $k$, the MSO-theory of $[1, \ldots k] \times {\cal N}$ is decidable. 
\end{proposition}

In ``monadic second-order transductions'' as developed by Courcelle (see \cite{CouEng12})  the operations 
of MSO-interpretations and of $k$-fold copying are combined into one. 

A third result needed in a later section is concerned with parity games over infinite game arenas. 
We refer to \cite{Tho97} for background. We consider a parity game graph as a structure 
$G = (V, P_0, P_1, E, C_1, \ldots , C_r)$ where $V$ is the (at most countable) set of vertices, $P_0$ and $P_1$ are 
unary predicates defining the partition of $V$ into the vertices of player 0 and player 1, respectively  (we use these 
names rather than Input and Output to be in accordance with the literature on parity games), 
$E$ is the edge relation, and the $C_i$ define a partition of $V$ where $v \in C_i$ means that 
vertex $v$ carries color (or priority) $i$. 

It is well-known that the parity game over $G$ is determined with positional winning strategies \cite{EmersonJutla91}.

In \cite{Wal02} it is shown that the winning regions of the two players are MSO-definable 
(by MSO-formulas $\varphi_j(x),$ for $j = 0,1$). Moreover, as we shall see, under certain conditions the 
standard proof of positional 
determinacy yields an extension of this result, namely that for each player there is 
an MSO-definable winning strategy on the respective winning region. The definition of a 
winning strategy is given by a formula $\psi_j(x,y)$ for the respective player $j$, 
 such that for each $u$ in the winning region of player $j$, there is exactly one 
vertex $v$ such that $(u,v)$ satisfies $\psi_j(x,y)$, where $v$ is the choice determined by the 
considered winning strategy of player $j$, from $u$. In order to guarantee this definability, 
we proceed in two steps: We show (in a later section) that for reachability games 
over the game arenas considered here, MSO-definable winning strategies exist, and then 
lift this result to parity games. (In a reachability game, a play is won by Output if 
it reaches a given MSO-defined target set at some point.) We settle the second step  
in the following poposition. 

\begin{proposition}\label{mso-strategy-parity}
Let $G $ be a parity game graph over $V$ with MSO-definable sets of priorities. 
The winning regions of the two players of the associated
parity game are MSO-definable. Moreover, if in each reachability game over $G$ with MSO-definable 
target set there is an MSO-definable positional winning strategy of the winner on his/her winning region, then this 
also holds for the considered parity game over $G$ .
\end{proposition}

The first part of the claim is shown in \cite{Wal02}. For the second, we proceed by induction on the 
number of colors (priorities) of the game graph under consideration, following the standard determinacy proof 
as given in \cite{Tho97}. If there is one color $r$ only, 
the claim is trivial (since player 0 wins by any choice if $r$ is even, and player 1 wins by any choice if $r$ is odd).
Assume now we have formulas $\varphi_i^k(x)$ defining the winning region $W_i^k$ of player $i$ in a game 
with $k$ colors, and formulas $\psi_i^k(x,y)$ defining a winning strategy of player $i$ over $W_i^k$ with $k$ colors. 
Let us treat the case where $k+1$ is even (the other works by exchanging the players). 
Using the formula $\varphi_0^{k+1}$ defining $W_0^{k+1}$ in the considered game with $k+1$ colors (known from 
\cite{Wal02}) we obtain $\psi_0^{k+1}(x,y)$ as follows. The winning region $W_0^{k+1}$ is composed of the 
attractor $A = A_0(C_{k+1} \cap W_0^{k+1})$ of player 0 and the complement of this set in $W_0^{k+1}$. This complement 
does not contain vertices with color $k+1$, so we have a formula $\varphi_0^k(x,y)$ defining a winning strategy 
for player 0 on this set. For the attractor, we note that it is MSO-definable by a formula $\varphi_A(x)$ saying 
``$x$ is in all sets $X$ containing  $C_{k+1} \cap W_0^{k+1}$ and satisfying the following closure properties'':
\[
\forall z \biggl[ \Bigl(z \in P_0 \wedge \exists z' \bigl(E(z,z') \wedge X(z')\bigr) \rightarrow z \in X\Bigr)
\wedge \Bigl(z \in P_1 \wedge \forall z' \bigl(E(z,z') \rightarrow X(z')\bigr) \rightarrow z \in X\Bigr) \biggr]
\]
We now invoke the assumption on the MSO-definability of the winning strategy of Output in the reachability game 
with target set $C_{k+1} \cap W_0^{k+1}$ over his winning region (which is $A$), say by the formula $\psi(x,y)$. 
We thus obtain an MSO-definable winning strategy of Output over $W_0^{k+1}$ by a formula saying 
\[
x \in  W_0^{k+1} \setminus A \rightarrow \varphi_0^k(x,y) \ \ \wedge \ \ x \in A \rightarrow \psi(x,y)
 \]

\section{Automata Models for Sequences of Natural Numbers}\label{sec-automata}

\subsection{\texorpdfstring{$\nats$-Memory Automata}{N-Memory Automata}}

In this section, we introduce \emph{$\nats$-memory automata}, which work on
sequences of natural numbers.
Such a sequence $\alpha = a_0a_1a_2\ldots \in \nats^\omega$ is represented by a
labeled grid as illustrated in Figure \ref{fig:grid}, where each number $a_i$ is
represented by a column:
In the $i$th column of the grid, the first $a_i$ nodes, starting from the
bottom, are labeled with $1$ and the remaining nodes are labeled with $\bot$.
For technical reasons, a node labeled with $\#$ is added at the bottom of every
column.

Formally, the \emph{grid representation} of a sequence
$\alpha = a_0a_1a_2\ldots \in \nats^\omega$ is a function
$g_\alpha\colon \nats \times \nats \to \{1,\bot,\#\}$ labeling the positions of
the grid with
\[
	g_\alpha(i,j) =
		\begin{cases}
			\#   & \text{if $i=0$,} \\
			1    & \text{if $1 \leq i \leq a_j$,} \\
			\bot & \text{if $i > a_j$.}
		\end{cases}
\]

An $\nats$-memory automaton can traverse a grid representation by moving up and
down within the current column or switching to the next column (but not the
previous one).
It has a finite set of states, but is additionally equipped with a
\emph{memory token}, which marks a row of the grid, and a
\emph{memory update token}, which can be placed by the automaton at the 
end of processing a column and which determines the position of the memory token 
on the next column.

For a formal definition, let $D = \{\uparrow,\downarrow,\rightarrow,\diamond\}$
be the set of possible actions of the automaton (move up, move down, switch to
next column, place memory update token).

\begin{definition}[$\nats$-Memory Automaton]
	An \emph{$\nats$-memory parity automaton} is a tuple
	$\aut{A} = (Q,q_0,\Delta,c)$ where $Q$ is a finite set of states,
	$q_0 \in Q$ is the initial state,
	$\Delta \subseteq Q \times \{1,\bot,\#\} \times \{0,1\}^2 \times Q \times D$
	is the transition relation with
	$\Delta \cap \Bigl( Q \times \{\#\} \times \{0,1\}^2 \times Q \times \{\downarrow\} \Bigr) = \emptyset$,
	and $c\colon Q \to \{0,\dotsc,m\}$ is a function assigning priorities to the
	states.
\end{definition}

We call $\aut{A}$ deterministic if for all
$\bigl(p,a,(b_1,b_2)\bigr) \in Q \times \{1,\bot,\#\} \times \{0,1\}^2$, there
is exactly one pair $(p,d) \in Q \times D$ such that
$\bigl(p,a,(b_1,b_2),p,d\bigr) \in \Delta$.

A \emph{configuration} of $\aut{A}$ is a tuple
$(q,h,v,i,j) \in Q \times \nats^4$, where $q$ is the current state, $h$ is the
horizontal position of the automaton (i.e., the current column), $v$ is the
vertical position (within the current column), and $i,j$ are the current positions of
the memory token and the memory update token, respectively, on the current column.
(If the memory update token is not yet placed, we assume position 0 for it by default.) 

A run of $\aut{A}$ on a sequence $\alpha \in \nats^\omega$ is an infinite
sequence $\pi = c_0 c_1 c_2 \ldots$ such that $c_0 = (q_0,0,0,0,0)$ is the
initial configuration and for every pair of consecutive configurations
$c_\ell=(q_\ell,h_\ell,v_\ell,i_\ell,j_\ell)$,
$c_{\ell+1}=(q_{\ell+1},h_{\ell+1},v_{\ell+1},i_{\ell+1},j_{\ell+1})$, one of the following holds:
\begin{itemize}
	\item $\bigl( q,g_\alpha(v_\ell,h_\ell),(b_1,b_2),q_{\ell+1},\uparrow \bigr) \in \Delta$, and $v_{\ell+1} = v_\ell + 1$, $h_{\ell+1} = h_\ell$, $i_{\ell+1} = i_\ell$, $j_{\ell+1} = j_\ell$, or
	\item $\bigl( q,g_\alpha(v_\ell,h_\ell),(b_1,b_2),q_{\ell+1},\downarrow \bigr) \in \Delta$, and $v_{\ell+1} = v_\ell - 1$, $h_{\ell+1} = h_\ell$, $i_{\ell+1} = i_\ell$, $j_{\ell+1} = j_\ell$, or
	\item $\bigl( q,g_\alpha(v_\ell,h_\ell),(b_1,b_2),q_{\ell+1},\rightarrow \bigr) \in \Delta$, and $v_{\ell+1} = v_\ell$, $h_{\ell+1} = h_\ell + 1$, $i_{\ell+1} = j_\ell$, $j_{\ell+1} = 0$, or
	\item $\bigl( q,g_\alpha(v_\ell,h_\ell),(b_1,b_2),q_{\ell+1},\diamond \bigr) \in \Delta$, and $v_{\ell+1} = v_\ell$, $h_{\ell+1} = h_\ell$, $i_{\ell+1} = i_\ell$, $j_{\ell+1} = v_\ell$,
\end{itemize}
where \quad
$\displaystyle b_1 =
	\begin{cases}
		1 & \text{if $i_\ell = v_\ell$,} \\
		0 & \text{otherwise}
	\end{cases}$
\quad and \quad
$b_2 =
	\begin{cases}
		1 & \text{if $j_\ell = v_\ell$,} \\
		0 & \text{otherwise}
	\end{cases}$
\\[1ex]
indicate whether the memory token and the memory update token, respectively,
are at the current vertical position $v_\ell$.

A run $\pi$ is accepting if $max\Bigl( \textit{Inf}\bigl( c(\pi) \bigr) \Bigr)$
is even.
A sequence $\alpha \in \nats^\omega$ is accepted by $\aut{A}$ if the run 
 of $\aut{A}$ on its grid representation $g_\alpha$ is accepting.

For example, the singleton language $\{ 1234\ldots \}$ is recognized by a deterministic
$\nats$-memory automaton that works as follows:
It checks that there is exactly one $1$ in the first column, and moves the
memory update token to the row $1$.
After switching to the next column, the memory token now marks row $1$.
The automaton goes up to that row and checks that there is exactly one $1$
above that position.
Then it moves the memory update token to the position above the memory token and
switches to the next column, and so on. The states assumed in this process have color $2$; 
once the checking process fails, color $1$ is assumed. A variation of this idea 
shows the recognizability of the $\omega$-language $\nats^* 1 \nats^* 2 \nats^* 3 \nats^* \ldots$.
 
The language $\{ \alpha \in \nats^\omega \mid \alpha \text{ is unbounded}\}$ is
recognized by a deterministic $\nats$-memory automaton that moves the memory
update token to the position of the topmost $1$ of the current column if that
position is higher than the current position of the memory token.
After any move of the memory update token, the automaton goes to a state with the
even priority $2$, otherwise to a state with priority $1$.

We give three further examples of languages recognized by $\nats$-memory
automata (without proof):
\begin{enumerate}
	\item $\{m_0 m_1 m_2 \ldots \mid m_{i+1} = m_i + 1 \ \text{ or } \ m_{i+1} = m_i - 1\}$
	\item $\{ m_0 m_1 m_2 \ldots \mid m_{i+1} \text{ even iff } m_i \text{ odd} \}$
	\item
		$\begin{alignedat}[t]{1}
			\{ m_0 m_1 m_2 \ldots  \mid
			& \; m_{2i+2} = m_{2i}+1 \text{ if } m_{2i+1} \text{ even} \\
			& \; m_{2i+2} = m_{2i}-1 \text{ if } m_{2i+1} \text{ odd} \;\; \}
		\end{alignedat}$
\end{enumerate}

Thus, $\nats$-memory automata can recognize some interesting $\omega$-languages over $\nats$.
The ability to compare successive (and also ``distant'') letters and to define properties of unboundedness 
seems to be a feature that is missing
in known models of automata over the alphabet $\nats$.  
 Let us 
note that in the context of temporal logic, a related idea appears in \cite{CarapelleFKL15}; 
however there equality and incremental change of values from $\nats$ is restriced to 
occurrences within a bounded (time-)interval -- so a language such as 
$\nats^* 1 \nats^* 2 \nats^* 3 \nats^* \ldots$ (as mentioned above) is not covered. 

An alternative version of $\nats$-memory automata can be defined by abstracting
from the steps within one column and representing the steps from one column to
the next one by MSO-formulas.
In this description we use the product structure $Q \times \structnats$ with 
domain $Q \times \nats$ as defined in Section \ref{mso-prerequisites}.


An automaton with this logical specification of the transitions, 
which we call \emph{MSO $\nats$-memory automaton}, is of the
form $\aut{A} = (Q,q_0,\bigl( \varphi((p,x),y,(q,z)) \bigr)_{p,q \in Q},c)$.
(This notation indicates a formula $\varphi(r,s,t)$ with $P_p(r)$ and $P_q(s)$.)
The following condition should be satisfied: 
Starting in state $p$ with memory token on position $i$, after processing the input
number $m$, the automaton will reach state $q$ with memory token on the new
position $j$ iff $Q \times \structnats \models \varphi[(p,i), m, (q,j)]$. \footnote{Strictly speaking, 
$m$ is not an element of $Q \times \nats$; by abuse of notation we write $m$ to denote 
the element $(q_0, m)$.}
We call such a step of the automaton a \emph{macro transition}.

If the MSO-$\nats$-memory automaton is deterministic (as in the present paper), then for every $(p,i)$ and
$m$, there is exactly one $(q,j)$ such that
$Q \times \structnats \models \varphi[(p,i),m,(q,j)]$.

\begin{proposition}
	For every $\nats$-memory automaton, an equivalent MSO-$\nats$-memory
	automaton can be constructed.
\end{proposition}

The proof is straightforward but tedious regarding the details. 
The idea is to describe the segments of a computation on a given column letter 
from one placement of the memory update token to the next. This computation 
segment can visit a given position of the given column only $\leq |Q|$ times; 
otherwise a repetition of configurations occurs and the computation does not 
terminate. Hence such a run segment can be described by an existential 
MSO-formula with $|Q|^2$ existential set quantifiers.
The processing of a column is a sequence of such computation segments, ending
at the point where the automaton switches to the next column, so it is captured
by the transitive closure of the segment computations. It is 
easy to express this invoking the definability of transitive closure in MSO. 

Furthermore, let us list some properties (not needed below, however) that are proved similarly to \cite{CST}.\footnote{A 
more detailed study of $\nats$-memory automata -- including a systematic
analysis of closure properties and the inequivalence between the deterministic
and the non-deterministic model -- is the subject of a forthcoming paper by
P.~Landwehr and the authors.}
\begin{remark}
	\begin{enumerate}
		\item The emptiness problem for $\nats$-memory automata over words from $\nats^*$  and the emptiness problem for $\nats$-memory parity automata over $\nats^\omega$ are decidable.
		\item This fails when the automata are equipped with two memory tokens 
		(and memory update tokens). 
	\end{enumerate}
\end{remark}
 
\subsection{\texorpdfstring{$\nats$-Memory Transducers}{N-Memory Transducers}}

We use deterministic $\nats$-memory automata to represent winning conditions in
Gale-Stewart games in the Baire space.
To represent strategies in such games, we introduce $\nats$-memory transducers,
which are defined in close analogy  to $\nats$-memory automata, with two modifications: 
Firstly, there is no priority function as used for the parity acceptance condition (since we are not dealing 
with infinite runs). Secondly, there is 
 an additional token, the \emph{output token}; 
 used to indicate a natural number that is produced as output after
reading a word of natural numbers as  given input sequence.

Thus, we define an extended set of actions
$\widehat{D} = \{\uparrow,\downarrow,\rightarrow,\diamond,\square\}$, and the
transition relation is now of the form
$\Delta \subseteq Q \times \{1,\bot,\#\} \times \{0,1\}^3 \times Q \times \widehat{D}$.
In a transition of the form $(p,a,(b_1,b_2,b_3),q,\square)$, the output token is
placed at the current vertical position.

We will only be interested in deterministic $\nats$-memory transducers, where
for all $\bigl(p,a,(b_1,b_2,b_3)\bigr) \in Q \times \{1,\bot,\#\} \times \{0,1\}^3$,
there is exactly one pair $(p,d) \in Q \times \widehat{D}$ such that
$\bigl(p,a,(b_1,b_2,b_3),p,d\bigr) \in \Delta$.

An $\nats$-memory transducer works like an $\nats$-memory automaton, but it
distinguishes between input and output columns.
After processing a given input column, it switches to an output column, which is
unlabeled except for the tokens (initially just the memory token).
The position of the output token upon moving to the next column 
then indicates the output number at that point.

By processing input and output columns in alternation, the transducer produces
an output sequence $\beta = e_0 e_1e_2e_3\ldots \in \nats^\omega$ for a given
input sequence $\alpha = a_0 a_1a_2a_3\ldots \in \nats^\omega$, yielding the 
play $a_0 e_0 a_1 e_1 a_3 \ldots$.

\section{Solving Games in the Baire Space}\label{solving-games}

Our aim here is to prove the following result:

\begin{theorem}
	For a Baire space game $\Gamma(L)$ where $L \subseteq \nats^\nats$ is 
	defined by a deterministic $\nats$-memory parity automaton $\mathcal{A}$, one can
	\begin{itemize}
		\item decide who wins $\Gamma(L)$, and
		\item construct a winning strategy for the winner realized by an
			$\nats$-memory transducer.
	\end{itemize}
\end{theorem}

In order to show the theorem, we proceed in two steps, following a pattern as
known from the classical solution of Church's Problem in the Cantor space. 

	\begin{enumerate}
		\item Convert the automaton into a parity game with designated start vertex.\\
			(In contrast to the classical setting, the game arena will be infinite here.)
		\item Solve the parity game (finding the winner and computing a memoryless winning strategy). 
	\end{enumerate}
In the first subsection we deal with the first step and the decision about the winner, in the subsequent subsection 
we present the construction of the desired transducer. 

\subsection{Deciding the Winner}

To transform the given deterministic $\nats$-memory automaton $\mathcal{A}$, recognizing $L \in \nats^\nats$, into a game arena,
we first construct an equivalent MSO-$\nats$-memory automaton $\mathcal{A}'$.
We assume that the state set $Q$ can be partitioned into sets $Q_0$ and $Q_1$
such that all macro transitions from $Q_0$ lead to $Q_1$ and vice versa.
This can always be achieved using two copies of the original state set.

Now we construct 
a game arena $G_\mathcal{A}$ with
domain $Q \times \nats $, the relations as defined in Section \ref{mso-prerequisites} for $Q \times \mathcal{N}$, and the 
additional edges
\begin{align*}
	(p,i) & \xrightarrow{m} (q,j)
\end{align*}
according to the macro transitions of $\mathcal{A}'$. In the following, we call a tuple $(p,i)$ as it occurs here 
a ``configuration''.

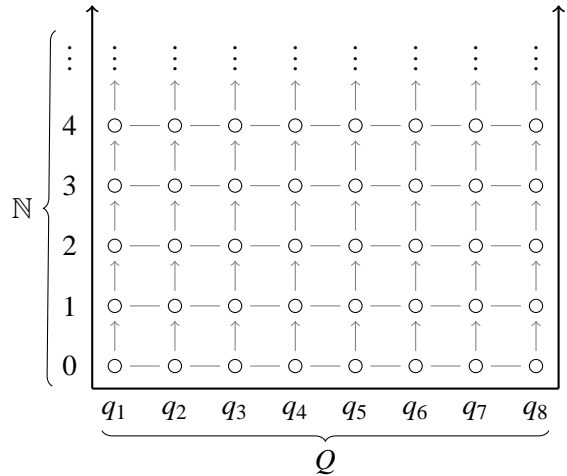
\begin{center}
\begin{tikzpicture}
	\tikzstyle{gamenode}+=[minimum size=5pt,inner sep=0pt]
	\tikzstyle{p0node}+=[gamenode,draw,circle]
	\tikzstyle{p1node}+=[gamenode,draw,rectangle]
	
	\newcommand{\pscale}{0.8}
	\newcommand{\dotsdist}{0.2}
	
	\tikzstyle{hedge}=[black!50, shorten >=3pt, shorten <=3pt]
	\tikzstyle{vedge}=[black!50, ->, shorten >=3pt, shorten <=3pt]
	
	\foreach \x in {0,...,7}
	\foreach \y in {0,...,4}
	{
		\node[p0node] (x\x y\y) at (\pscale*\x,\pscale*\y) {};
	}
	
	
	\foreach \x in {0,...,7}
	{
		\node[p0node,draw=none] (x\x y5) at (\pscale*\x,\pscale*5) {};
		\node at (\pscale*\x,\pscale*5+\dotsdist) {\vdots};
	}

	\foreach \x in {0,...,6}
	\foreach \y in {0,...,4}
	{
		\pgfmathtruncatemacro{\sx}{\x + 1}
		\draw[hedge] (x\x y\y) -- (x\sx y\y);
	}

	\foreach \x in {0,...,7}
	\foreach \y in {0,...,4}
	{
		\pgfmathtruncatemacro{\sy}{\y + 1}
		\draw[vedge] (x\x y\y) -- (x\x y\sy);
	}
	
	\newcommand{\axisdist}{0.3}
	
	\coordinate (southwest) at (-\axisdist,-\axisdist);
	\coordinate (northwest) at (-\axisdist,6*\pscale);
	\coordinate (southeast) at (7*\pscale+\axisdist,-\axisdist);
	\coordinate (northeast) at (7*\pscale+\axisdist,6*\pscale);
	
	\draw[thick] (southwest) edge[->] (northwest);
	\draw[thick] (southwest) edge[-] (southeast);
	\draw[thick] (southeast) edge[->] (northeast);
	
	\foreach \x in {0,...,7}
	{
		\pgfmathsetmacro{\xinc}{\x+1}
		\node (xlabel\x) at (\pscale*\x,-2*\axisdist) {$q_{\pgfmathprintnumber[int trunc]{\xinc}}$};
	}
	\draw[decorate,decoration={brace,amplitude=4pt,mirror},yshift=-0.4pt] ([xshift=4pt]xlabel0.south west) -- ([xshift=-4pt]xlabel7.south east) node[midway,yshift=-12pt] {$Q$};
	
	\foreach \y in {0,...,4}
	{
		\node (ylabel\y) at (-2*\axisdist,\pscale*\y) {$\y$};
	}
	\node (ylabeldots) at (-2*\axisdist,\pscale*5+\dotsdist) {\vdots};
	\draw[decorate,decoration={brace,amplitude=4pt,mirror},yshift=-0.4pt] ([yshift=-4pt]ylabeldots.north west) -- (ylabel0.south west) node[midway,xshift=-12pt] {$\nats$};
\end{tikzpicture}
\end{center}

\begin{lemma}
	$G_\mathcal{A}$ is MSO-interpretable in $Q \times {\cal N}$ . 
\end{lemma}

The proof is straightforward by describing the edge relations of 
$G_\mathcal{A}$ in $Q \times {\cal N}$. 
Thus we obtain the following proposition.

\begin{proposition}

The MSO-theory of $G_\mathcal{A}$ is decidable.

\end{proposition}

We now can decide the winner of $\Gamma(L)$.  For this we use 
the first claim of Proposition \ref{mso-strategy-parity} (Section \ref{mso-prerequisites}):
Describe the initial vertex $(q_0,0)$ of $G_\mathcal{A}$ by a formula
$\psi_\text{init}(x)$, and let $\varphi_\text{Out}(x)$ be a formula defining the winning region
of Player Output. We check whether
\[
	G_\mathcal{A} \models \exists x \bigl( \psi_\text{init}(x) \land \varphi_\text{Out}(x) \bigr)
\]

\subsection{Constructing a Transducer}

We treat here the case that the winner is Player Output. We first want to apply Proposition \ref{mso-strategy-parity} (Section \ref{mso-prerequisites}). 
So we have to show that the assumption of Proposition \ref{mso-strategy-parity} regarding reachability games holds for the games considered 
here, namely that an MSO-definable winning strategy for Output (over his winning region) exists for a reachability game 
over $G_\mathcal{A}$ with MSO-definable target set. 
Then, applying Proposition \ref{mso-strategy-parity}, we know that in the parity game over  $G_\mathcal{A}$, an MSO-definable 
winning strategy exists for Player Output on his winning region. In the second step we use this fact to  
obtain the desired transducer. 

\subsubsection{MSO-Definable Winning Strategies in Reachability Games}

We  show the following, referring to the game arena $G_\mathcal{A}$ introduced above.

\begin{proposition}\label{mso-strategy-reach}
In every reachability game over $G_\mathcal{A}$ with an MSO-definable target set $F$, Player Output has an MSO-definable 
positional winning strategy on his winning region. 
\end{proposition}

We show this claim by a transformation of the reachability game over $G_\mathcal{A}$ with target set 
$F$ into a pushdown reachability game $\mathcal{P}$ over an extended domain $P \times G_\mathcal{A}$ for some finite $P$. 
A transition $(p,i) \rightarrow (q,j)$ in $G_\mathcal{A}$ (via some input number $m$) will be dissolved into a sequence of steps 
over the pushdown arena $\mathcal{P}$, proceeding from stack content $\# 1^i$ to stack content $\# 1^j$ in steps each of 
which changes the stack only by 1. Some complications arise from the fact that a transition from $(p,i)$ to $(q,j)$ 
depends on an input value $m$ from the infinite domain $\nats$. As we shall see, we can handle this using finite information 
about $m$ when the target value $j$ is ``near'' to $0$ or $i$; otherwise the target value $j$ will be ``near'' to $m$, and 
the stack will be changed accordingly.

As a preparation we need an obvious fact on the behaviour of the deterministic automaton $\mathcal{A}$: 

\begin{lemma}
There is a bound $B$ such that from configuration $(p,i)$ with input $m$, the automaton $\mathcal{A}$ will 
reach an  exit configuration $(q,j)$ where the distance of $j$ to $0$ or $i$ or $m$ is bounded by $B$. 
\end{lemma}

The lemma is clear by the fact that between the marked positions $0, i, m$ in a column the automaton $\mathcal{A}$ is 
processing one-letter input words. On such words of sufficiently large length $B$, the automaton $\mathcal{A}$ will assume 
a periodic behaviour and hence would violate the condition that a unique value $j$ is reached upon termination. 

\bigskip

According to the lemma, the configuration upon leaving a column can be
represented by a tuple
$(q,t,k) \in Q \times \{\text{``0''},\text{``I''},\text{``M''}\} \times \{-B,\dotsc,B\}$,
which we call an \emph{exit combination}.
For example, the tuple  $(q,\text{``I''},2)$ would indicate that the column is
left in state $q$ with the memory update token on position $i+2$.
Note that the set $E$ of exit combinations is finite.

A second remark refers to the 
periodic behaviour of the deterministic automaton $\mathcal{A}$ on words over a singleton alphabet. 
Such word segments occur between 
the positions 0, the memory token position $i$, and the input position $m$. The states assumed 
by $\mathcal{A}$ occur periodically. There is a  finite prefix length $\ell_0$ 
and a period length $\ell$ (which can be taken as $|Q|!$) 
such that given any starting state $p$ at position $i$, the state of $\mathcal{A}$ at 
position $i - k$ or $i + k$ is fixed by the number in $[0, \ell_0 + \ell]$ which is identical to $k$ when $k \leq  \ell_0$ 
or otherwise in $[\ell_0 +1 , \ell_0 + \ell]$ and with same remainder modulo $\ell$ as $k$. 
Call this number the ``$(\ell_0, \ell)$-status of $k$'' (or just \emph{status} of $k$).

Note that for any $p \in Q$, $i,m \in \nats$, the corresponding exit combination
is determined by the status of $i$, the status of $m$, the status of $|i-m|$, and whether
$i<m$ (we refer to the last three items as the \emph{relative status} of $m$ with
respect to $i$). Writing $S$ for the set of individual status informations, and ${0,1}$ for the information 
whether $i < m$ or not, we obtain a finite (and effectively computable) 
relation $R \subseteq S^3 \times \{0,1\} \times E$ consisting of those 
tuples where the last component is determined by the first four components.

\bigskip

We now give a sketch of the proof of Proposition \ref{mso-strategy-reach}.
We define a pushdown arena $\mathcal{P}$ where, intuitively, the height of
the stack indicates the current position of the memory token.
The control states of the pushdown system indicate the current state $p$ of
$\mathcal{A}$ and also the status of the current stack height.

Consider a configuration of the pushdown system where the state of $\mathcal{A}$
is $p$ and the height of the stack, representing the memory token position, is
$i$ (and its status is stored in the control state).
The current player, say Output, can now choose a tuple $r \in R$
where the first component of $r$ is the status of $i$.
This amounts to a decision about the number $m$ that Player Output wants to play
in the original game: it fixes the relative status of $m$ with respect to $i$
(and thus the exit combination representing the behavior of
$\mathcal{A}$ on a column of height $m$).

In the following steps of the pushdown game, Player Output will modify the stack
content to represent the new memory token position $j$ according to the exit
combination $e$ that is determined by his choice of $r$.
If $e$ is of the form $(q,\text{``0''},k)$, he can empty the stack and
then increase its height to $k$.
For a combination $e=(q,\text{``I''},k)$, the height of the stack (currently
representing $i$) is increased/decreased by $k$.
If $e$ is of the form $(q,\text{``M''},k)$, the player can either
increase of decrease the height of the stack step by step.
While the stack is modified, the relative status of the current stack height
with respect to $i$ is tracked in the control state of the
pushdown system.
Whenever the current height of the stack is a number $m$ with the previously chosen
relative status (given by $r$), the player can finally increase/decrease the height by
$k$, which determines the new memory token position.

Now we can apply the fact that attractor strategies in pushdown reachability games are
definable by finite automata (see \cite{CarayolH14}) -- and hence in MSO-logic. \footnote{In \cite{CarayolH14},
also  parity games are mentioned; for easier presentation we consider reachability games and
apply Proposition \ref{mso-strategy-parity} for the step to parity games.}
\begin{proposition}
    Positional winning strategies in pushdown reachability games with MSO-definable target set can be
    implemented by deterministic finite automata reading a given pushdown
    configuration and yielding as output the pushdown rule to be applied next.
\end{proposition}

In this result, the choice of the next move is fixed by the name $h$ of the pushdown rule
to be applied. In MSO-logic, we obtain thus formulas
$\psi_h(x)$ that are true if for position $x = (p,i)$ the rule to be applied is $h$. It is easy to
transform these MSO-formulas into  a single MSO-formula $\chi(x,y)$ which fixes $y$ as the element reached from $x$
by applying the unique rule $h$ where $\psi_h(x)$ is true.

In the last step, we have to combine the finitely many steps of a player in $\mathcal{P}$
forming altoghether a macro transition of $\mathcal{A}$ into a single step,
and we have to transfer the MSO-definability of the strategy from the arena
$P \times G_\mathcal{A}$ (i.e., $P \times Q \times \structnats$) of the pushdown
game to the structure  $Q \times \structnats$.

To combine the intermediate steps forming a macro transition, we apply the
(MSO-definable) transitive closure to the strategy formula $\chi(x,y)$ for the
player under consideration, with the requirement that an exit configuration is
finally reached, yielding another MSO-formula $\chi'(x,y)$.

To obtain an MSO-definable strategy over the original arena $Q \times \structnats$,
it suffices to note that the finitely many
tuples of $S^3 \times \{0,1\} \times E$ can be coded in a finite label alphabet
and that the status information of numbers is definable in MSO-logic.

\subsubsection{From MSO-Definability of Strategies to Transducers}

\begin{proposition}\label{prop:transducer}
	Given an MSO-definable winning strategy of Player Output in the parity game
	on $G_\mathcal{A}$, there is an $\nats$-memory transducer
	realizing a winning strategy in $\Gamma(L(\mathcal{A}))$.
\end{proposition}

Assume Player Output wins $\Gamma(L(\mathcal{A}))$.
By Proposition \ref{mso-strategy-parity}, we have an MSO-formula $\varphi(x,y)$ 
defining a winning strategy on his winning region $W_\text{Out}$ of $G_\mathcal{A}$.
For the construction of the transducer, we will use the following lemma.

\begin{lemma}\label{lemma:checkautomaton}
For a given MSO-formula $\varphi(x,y)$ over $Q \times \structnats$
and given $p,q \in Q$, we can construct a deterministic finite automaton
$\aut{C}^\varphi_{pq}$, whose input is a column (i.e., a word) that is
unlabeled except for tokens at positions $i,j$
(memory token and memory update token), that terminates and that accepts iff
$Q \times \structnats \models \varphi[(p,i),(q,j)]$.
\end{lemma}

This automaton is obtained as follows:
For a formula $\varphi(x,y)$ over $Q \times \structnats$, we can construct
corresponding formulas $\varphi'_{pq}(x',y')$ over  ${\cal N}$ such that
$\structnats \models \varphi'_{pq}[i,j]$ iff
$Q \times \structnats \models \varphi[(p,i),(q,j)]$.
To obtain such a formula, each second-order variable $X$ in $\varphi$ is replaced
by a $|Q|$-tuple of second-order variables $(X_q)_{q \in Q}$ (see \cite{BCL07}).

Then the resulting MSO-formula can be translated into an equivalent Büchi
automaton, which in turn can be converted into an NFA that accepts or rejects
immediately after the last of the two tokens in the column has been read, depending on
whether the Büchi automaton can reach an accepting loop on the unlabeled rest of
the column.
This NFA can be determinized, yielding the desired automaton
$\aut{C}^\varphi_{pq}$.

Using Lemma \ref{lemma:checkautomaton}, we can now construct the transducer as
claimed in Proposition \ref{prop:transducer}. Note that the formula $\varphi(x,y)$ defining a 
winning strategy fixes a unique update for a configuration $(p,i)$ to a configuration $(q,j)$. 
For the output of the transducer we have to find a number $m$ such that $(p,i) \xrightarrow{m} (q,j)$
is a possible transition in the game graph $G_\mathcal{A}$. 
The transducer will go through the possible values of $m$, by placing the output token successively on 
position $0,1,2, \ldots$. In each case, say with the output token on position $m$, 
it works like $\aut{A}$ to find from start configuration $(p,i)$ the new configuration $(q,j)$. 
Now   $\aut{C}^\varphi_{pq}$ is used to check whether the move to $(q,j)$ is in accordance  
with the winning strategy. If this is the case, the current value of $m$ is the desired output. 

In more detail: Assume that the transducer has processed an input column and has just switched
to the subsequent output column, in state $p$ and with the memory token at
position $i$.
Starting with the output token on position $0$, the transducer now proceeds as
follows:
It simulates the $\nats$-memory automaton $\aut{A}$ (including the placements of
the memory update token) on the column $\#1^m\bot^\omega$, where $m$ (initially $m = 0$) is
the current position of the output token.

At some point, $\aut{A}$ would switch to the next column.
Let $j$ be the position of the memory update token and let $q$ be the
state of $\aut{A}$ at that point.
The transducer now invokes the automaton $\aut{C}^\varphi_{pq}$ described in
Lemma \ref{lemma:checkautomaton} to check whether $(q,j)$ is the correct target
position according to the strategy given by $\varphi(x,y)$.
If this is the case (i.e., $\aut{C}^\varphi_{pq}$ accepts) then the transducer
terminates processing the current column (and moves to the next input column).
Otherwise, it moves the output token one position upwards and repeats the steps
above.
At some point, the correct target configuration $(q,j)$ will be found, so the transducer will
eventually produce the desired output number. 

\section{Summary and Perspectives}

We have introduced $\nats$-memory automata as a natural model of automata over the infinite alphabet $\nats$, 
and in this framework we have obtained an algorithmic solution of Church's synthesis problem. It seems to be 
the first algorithmic solvability result on games in the Baire space. 

Let us address some open issues: 

\begin{enumerate}
\item Find a more direct construction for the decision of the winner and the winning strategy. We have invoked 
decidability results on MSO-theories. 

\item Related to the first issue, a complexity analysis should be supplied -- this is missing in the present paper.

\item One may wonder whether a logical framework of game specifications can be developed, replacing the 
presentation in terms of $\nats$-memory automata. This, however, seems difficult, since the class of $\omega$-languages 
recognized by $\nats$-memory automata has only poor logical closure properties (for instance, already closure 
under intersection fails). 

\item How can one strengthen the model of $\nats$-memory automaton, still keeping decidability results as needed 
to obtain an algorithmic solution of Church's synthesis problem?

\item Replace plays over $\nats$ by plays over $\Sigma^*$ for finite $\Sigma$.

\item A related problem is to find such results relying on decidability of the MSO-theory of 
the infinite binary tree rather than of $(\nats, \mathit{Succ})$. 
\end{enumerate}

\bibliographystyle{eptcs}
\bibliography{bibliography}
\end{document}